\documentclass[12 pt, amssymb,prb,showpacs]{revtex4}
\usepackage{graphicx}
\usepackage{epsfig}
\usepackage{dcolumn}
\begin{document}
%\title{Preparing a paper using \LaTeXe\ for publication in \jpcs}

%\author{Jacky Mucklow}

%\address{Production Editor, \jpcs, \iopp, Dirac House, Temple Back, Bristol BS1~6BE, UK}

%\documentclass[12pt]{iopart}
%\newcommand{\gguide}{{\it Preparing graphics for IOP journals}}

%\usepackage{iopams}
%\usepackage{epsfig}
%\usepackage{dcolumn}
%\usepackage{amsmath}
\hyphenation{semi-con-ductor}
%\begin{document}

\title{\bf Influence of linearly polarized radiation on magnetoresistance
in irradiated two-dimensional electron systems}

\author{Jes\'us I\~narrea}
 \address{$^1$Escuela Polit\'ecnica
Superior,Universidad Carlos III,Leganes,Madrid,28911,Spain}
%\ead{jinarrea@fis.uc3m.es}

\date{\today}

 % For example: 71.20.Ps
%\date{\today}

%\pacs{73.40.-c}{First pacs description} \pacs{73.50.-h}{Second pacs
%description} \pacs{78.67.-n}{Third pacs description}

%%%%%%%%%%%%%%%%%%%%%%%%%%%%%%%%%%%%%%%%%%%%%%%%%%%%%%%%%%%%%%%%%%
%%%%%%%%%%%%
\begin{abstract}
We study the influence of the polarization angle of linear
radiation on the radiation-induced
magnetoresistance oscillations in two-dimensional electron systems, and
examine the polarization immunity on the temperature
and quality of the sample.
We have applied  the radiation-driven electron orbits model obtaining that the magnetoresistance is  affected
by the orientation of the electric field of linearly polarized
radiation when dealing with high quality samples and low temperatures. Yet, for lower quality  samples and higher
temperature
we recover polarization immunity in the radiation driven magnetoresistance oscillations.
This could be of interest
for future photoelectronics in high quality  mesoscopic devices.
\end{abstract}

%%%%%%%%%%%%%%%%%%%%%%%%%%%%%%%%%%%%%%%%%%%%%%%%%%%%%%%%%%%%%%%%%%

%%%%%%%%%%%%
\maketitle
\newpage

%\section{Introduction}
Important and unusual properties
have been discovered when two-dimensional electron
systems (2DES) are subjected to external AC or DC fields\cite{ina0}. We can
highlight the recently measured effect of microwave-induced
resistance oscillations (MIRO)\cite{mani,zudov} and zero
resistance states (ZRS)\cite{mani,zudov}. They are obtained
when two-dimensional electron systems, in high
mobility samples at low temperature ($\sim 1K$), are subjected to a perpendicular magnetic field ($B$) and
radiation (microwave (MW) band) simultaneously. In these experiments, for an increasing
radiation power ($P$), one first obtains longitudinal
magnetoresistance  oscillations
  which evolve
into zero resistance states  at high enough $P$.
Early experiments obtained similar MIRO when radiation
is linearly polarized either in the transport direction
or perpendicular to it\cite{maniapl}. A recent study utilizing a new
setup has shown, however, a marked linear polarization
sensitivity\cite{manipolar,jesuspolar}
Other
experiments obtained that the  magnetoresistivity response
$\rho_{xx}$ is independent of the direction of the circular
polarization of radiation\cite{smet}. Some theoretical
contributions
\cite{ina2,girvin,dietel,lei,rivera,shi,islamov}
have been presented trying to explain
such striking effects but to date, there is
no consensus on the physical origin . In this paper we present a theoretical model
to treat the physical problem of a 2DES subjected simultaneously to
a static, moderate and uniform magnetic field ($B$) and
radiation \cite{ina2,kerner} linearly polarized at any
angle. We study how the polarization immunity first detected
in previous experiments can be altered when dealing with
higher quality samples and lower temperatures. We propose that
under such conditions, radiation-driven magnetoresistance can be affected
depending on the different angles of linearly polarized radiation.
The theory and results presented here are important to
understand, more in deep, the coupling between matter and radiation and
in the presence of static $B$ and
it could be of interest for the development
of future photoelectronics in high quality mesoscopic devices.

%\begin{figure}
%\centering\epsfxsize=3.0in \epsfysize=3.7in
%\epsffile{JOFAgrafico1.eps} \caption{Schematic diagram showing a
%2DEG illuminated by linearly polarized MW radiation at different
%polarization angles ($\alpha$). a) $\alpha=0$: the MW electric
%field, is aligned to the transport direction ($x$ direction). b)
%$0<\alpha<\pi/2$. c) $\alpha= \pi/2$: the MW electric field is
%perpendicular to $x$.}
%\end{figure}
%\section{Theoretical model}
We consider a 2DES ($x-y$ plane) subjected to a perpendicular ($z$
direction), static $B$, a DC electric field $E_{dc}$, responsible of
the electron transport through the sample ($x$ direction).
The system is also subjected to linearly polarized MW
radiation. The electric field $\overrightarrow{E}(t)$ of MW can be
in different polarization angles ($\alpha$) (see Fig. 1). This field
is given by:
%\begin{equation}
$\overrightarrow{E}(t)=(E_{0x} \overrightarrow{i}+E_{0y}
\overrightarrow{j})\cos wt$
%\end{equation}
where $E_{0x}$, $E_{0y}$ are the amplitudes  of the MW field and $w$
the frequency. Thus, $\alpha$ is given by
%\begin{equation}
$\tan \alpha= \frac{E_{0y}}{E_{0x}} $.
%\end{equation}
Considering the symmetric gauge for the vector potential of $B$,
$(\overrightarrow{A_{B}}=-\frac{1}{2}\overrightarrow{r}\times
\overrightarrow{B})$ the corresponding Hamiltonian reads:
\begin{eqnarray}
H&=&\frac{P_{x}^{2}+P_{y}^{2}}{2m^{*}}+\frac{w_{c}}{2}L_{z}+\frac{1}{2}m^{*}\left[\frac{w_{c}}{2}\right]^{2}
\left[(x-X)^{2}+y^{2}\right]\nonumber \\
&&-\frac{e^{2}E_{dc}^{2}}{2m^{*}\left[\frac{w_{c}}{2}\right]^{2}}\nonumber\\
&&-eE_{0x}(x-X)
\cos wt -eyE_{0y}\cos wt -eE_{0}X \cos wt\nonumber\\
\end{eqnarray}
 $X=\frac{eE_{dc}}{m^{*}(w_{c}/2)^{2}}$ is the center of the orbit for
the electron cycloidal motion, $w_{c}$ is the cyclotron frequency,
$L_{z}$ is the z-component of the electron total angular momentum.
Remarkably, the exact analytic solution of the Hamiltonian $H$ can be obtained\cite{ina2,kerner}:
\begin{eqnarray}
\Psi(x,y,t)\propto\phi_{N}\left[(x-X-a(t)),(y-b(t)),t\right]
%&&\times  \exp \frac{i}{\hbar} \left[m^{*}\left(\frac{d a(t)}{dt}x+\frac{d b(t)}{dt}y\right)+
%\frac{m^{*}w_{c}(b(t)x-a(t)y)}{2}
%-\int_{0}^{t} {\it L} dt'\right]\nonumber\\
%&&\times \sum_{p=-\infty}^{\infty} J_{p}(A_{N}) e^{ipwt}
\end{eqnarray}
%where  $J_{p}$ are Bessel functions with arguments,
%$A_{N}$\cite{ina3},   $\phi_{N}$  the well-known Fock-Darwin
%states\cite{fock} and $L$ the classical Lagrangian.
%This is the exact form of the analytical solution of a
%two-dimensional quantum oscillator under an external time dependent
%force.
where $\phi_{N}$ are the well-known Fock-Darwin states\cite{fock}.
The most important result with a deep physical meaning is
that, apart from phase factors, the wave function  $\Psi$ is proportional to
a Fock-Darwin state where the guiding center of the 2D
oscillator is displaced by $a(t)$ in the $x$ direction and $b(t)$ in
the y direction, i.e., describing an approximately  circular motion in the
$x-y$ plane. This motion is reflected in the x direction as
harmonic oscillatory with the MW frequency $w$.
The magnitude and nature of this displacement and
its physical effects will depend on the type of the time-dependent
force.

The expression for $a(t)$ and $b(t)$ can be obtained for a definite
polarization angle of the MW electric field with respect to
the x direction, $\alpha$,\cite{ina2}:
\large
\begin{eqnarray}
a(t)&=&\frac{eE_{0}\cos
wt}{m^{*}\sqrt{\frac{w^{2}(w_{c}^{2}-w^{2})^{2}}{w^{2}\cos^{2}
\alpha + w_{c}^{2}\sin^{2} \alpha}+\gamma^{4}} }   =A \cos wt\nonumber\\
\\
b(t)&=&\frac{e E_{0}\sin wt}{m^{*}\sqrt{\frac{w_{c}^{2}(w_{c}^{2}-w^{2})^{2}}{w^{2}\cos^{2}
\alpha + w_{c}^{2}\sin^{2} \alpha}+\gamma^{4}} }\nonumber\\
\end{eqnarray}
\normalsize
$\gamma$ is a material and sample-dependent damping factor which
dramatically affects the movement of the MW-driven electronic
orbits\cite{ina3}. Along with this movement interactions occur
between electrons and lattice ions, yielding acoustic phonons and
producing a damping effect in the back and forth electronic orbits motion. This parameter
$\gamma$ is going to play a crucial role in how the polarization angle affects
the magnetoresistance.
%\begin{figure}
%\centering\epsfxsize=3.5in \epsfysize=5.0in
%\epsffile{JOFAgrafico2.eps} \caption{ a) Calculated magnetoresistivity
%$\rho_{xx}$ as a function of $B$ for linearly polarized radiation
%and different polarization angles $\alpha$ regarding the transport
%direction ($x$ direction). In b) curves are shifted for
%clarity. The $\rho_{xx}$ polarization immunity can be clearly
%observed, especially for $B$ below the cyclotron resonance (see
%vertical dashed line in the top panel). T=1.5K.}
%\end{figure}
Now we introduce the scattering suffered by the electrons due to
charged impurities randomly distributed in the
sample\cite{ridley,ina2}. Firstly we calculate the electron-charged
impurity scattering rate $1/\tau$, and secondly we find the average
effective distance advanced by the electron in every scattering
jump:
%\begin{equation}
$\Delta X^{MW}=\Delta X^{0}+ A \cos w\tau$
%\end{equation}
where $\Delta X^{0}$ is the effective distance advanced when there
is no MW field present. Finally the longitudinal conductivity
$\sigma_{xx}$ can be calculated: $\sigma_{xx}\propto \int dE
\frac{\Delta X^{MW}}{\tau}$,  being $E$ is the energy. To obtain
$\rho_{xx}$ we use the usual tensor relationships
$\rho_{xx}=\frac{\sigma_{xx}}{\sigma_{xx}^{2}+\sigma_{xy}^{2}}
\simeq\frac{\sigma_{xx}}{\sigma_{xy}^{2}}$, where
$\sigma_{xy}\simeq\frac{n_{i}e}{B}$ and $\sigma_{xx}\ll\sigma_{xy}$.
%\begin{figure}
%\centering\epsfxsize=3.5in \epsfysize=3.8in
%\epsffile{JOFAgrafico3.eps} \caption{Calculated magnetoresistivity
%$\rho_{xx}$ as a function of $B$ for linearly polarized radiation
%and polarization angles $\alpha=0º$ and $\alpha=90º$ regarding the transport
%direction ($x$ direction). These results correspond to
%a scenario of high quality sample and lower temperature ($T=0.7K$).
%We observe how the $\rho_{xx}$ polarization immunity can be clearly
%altered in function of the angle. }
%\end{figure}

%\section{Results}
According to our model, the $\rho_{xx}$ response under MW excitation
is governed by the term $A \cos w\tau$: $\rho_{xx} \propto A \cos
w\tau$. Therefore we would expect different results of $\rho_{xx}$
depending on the angle $\alpha$ since $\rho_{xx}$ depends on
$\alpha$ through  the amplitude $A$ (see equation 3). However if the damping factor
$\gamma$ is larger than the MW frequency, $\gamma>w$, $\gamma$ would
become the leading term in the denominator of $A$. In this
situation, $\gamma$ is able to quench the influence of the other
terms. Therefore similar values are obtained for the amplitude of
the orbit center for different $\alpha$. For GaAs and typical
experimental MW frequencies\cite{mani,ina3,mani2}, a value for
$\gamma$ about $1-2\times10^{12}s^{-1}$ is enough to obtain a
similar $\rho_{xx}$ response irrespective of the value of $\alpha$.
%As we said previously, the first important result we obtain is that,
%apart from phase factors, the wave function for $H$ is the same as a
%Fock-Darwin state where the center of the electron orbits performs a
%circular motion in the xy plane with frequency $w_{c}$, given by
%$a^{2}(t)+b^{2}(t)=A^{2}$. In the x axis this circular motion is
%reflected as  harmonic oscillatory with the same frequency $w_{c}$.
%This MW-induced oscillating motion affects dramatically the way in
%which electrons in their orbits interacts with scatterers compared
%to the dark case. The interplay between the {\it transformed}  exact
%states (see eq. 48) and the charged impurity scattering between
%these states, is the origin of the non-linear effects observed in
%$\rho_{xx}$, i.e., MIRO´s and ZRS. A more extensive explanation of
%the physical origin of MIRO´s and ZRS can be obtained in
%ref.\cite{ina,ina10}.
In Fig.2 we show the calculated $\rho_{xx}$ as a function of $B$ for
linearly polarized radiation and for different polarization angles.
MW frequency $w=90$ GHz.  $\rho_{xx}$ response is practically immune
to the polarization angle, specially for $B$ below cyclotron
resonance (see vertical dashed line, Fig. 2a).
\\
\\
In the same way we would be able to obtain an entirely different
scenario lowering the parameter $\gamma$ and making the frequency
term the leading part in the denominator of $A$. We have to remember
that the polarization angle $\alpha$ is part of the frequency term in $A$ and
in this new scenario  different values of $\alpha$ will be reflected in
the frequency term and eventually in the obtained magnetoresistance. We have developed
a microscopical model for $\gamma$\cite{ina2,ina5} that summarizes
the interaction of electron with lattice ions giving rise to acoustic
phonons as:
\\
\\
\\
\begin{eqnarray}
\gamma&=&\frac{2\Xi^{2}m^{*}k_{B}T}{v_{s}^{2}\rho \pi \hbar^{3}<z>}\times \nonumber\\
&&\Bigg\{1+2\sum_{s=1}^{\infty}exp\left[-\frac{\pi\Gamma s}{\hbar w_{c}}\right]\cos\left[\frac{2\pi s \hbar w_{ac}}{\hbar w_{c}} \right]\Bigg\}\nonumber\\
\end{eqnarray}
where $\Xi$ is the acoustic deformation potential, $\rho$ the
mass density, $k_{B} $ the Boltzman constant, $T$ the temperature, $v{s}$ the sound velocity and $<z>$ is the effective
layer thickness. When the total sum inside brackets is carried out we obtain\cite{grads} after some
algebra:
\begin{equation}
\large
\gamma=\frac{2\Xi^{2}m^{*}k_{B}T}{v_{s}^{2}\rho \pi \hbar^{3}<z>}\left( \frac{1-e^{\frac{-\pi\Gamma}{\hbar w_{c}}}}{1+e^{\frac{-\pi\Gamma}{\hbar w_{c}}}}\right)
\end{equation}
\\
\\
The latter expression expresses the physical fact that
 for high quality samples  the
Landau level width gets smaller and in an inelastic scattering event (phonon emission)
is increasingly difficult to find a final Landau state where to get to.
Then, the term inside brackets decreases, when compared to a lower quality samples. In other words, for high quality samples , the
damping parameter $\gamma$ will decrease, making increasingly difficult
the damping by acoustic phonon emission and the release of the absorbed energy to the lattice.
Therefore, we have
a {\it bottleneck effect} for the emission of acoustic phonons.
On the other hand, $\gamma$ is linear with $T$ and the use of
lower temperatures will make smaller the damping parameter too.
Summarizing, the joint effect of high quality samples plus lower
temperatures makes  possible
to reach a situation where the frequency term is more important than
the damping term, ($\gamma < w$) making visible a
clear effect of the polarization angle on the magnetoresistance.
For GaAs high quality experimental parameters and $T<1 K$, we obtain that $\gamma\simeq 1-2 \times
10^{11} s^{-1}$. In Fig. 3 we present calculated magnetoresistivity
 as a function of $B$ for linearly polarized radiation
with polarization angles $\alpha=0º$ and $\alpha=90º$ regarding the transport
direction ($x$ direction). These results correspond to
a scenario of high quality sample and lower temperature ($T=0.7K$).
We observe that  the $\rho_{xx}$ polarization immunity can be clearly
altered in function of the angle.  This theoretical results have been confirmed
by recent experiments on linearly polarized MW-induced
resistance oscillations\cite{manipolar,jesuspolar}.
We consider that the results presented here are important to
understand the coupling between matter and radiation when an static $B$
is present and
it could be of interest for the development
of future photoelectronic   mesoscopic devices.

This work is supported by the MCYT  grant
MAT2011-24331 (Spain) and by the ITN Grant 234970 (EU).

%\section{References}

\newpage
\clearpage

Figure 1 caption:
Schematic diagram showing a
2DEG illuminated by linearly polarized MW radiation at different
polarization angles ($\alpha$). a) $\alpha=0$: the MW electric
field, is aligned to the transport direction ($x$ direction).  $\overrightarrow{E_{dc}}$ is the 
applied dc electric field responsible of the current. b)
$0<\alpha<\pi/2$. c) $\alpha= \pi/2$: the MW electric field is
perpendicular to $x$.
\newline

Figure 2 caption: a)
a) Calculated magnetoresistivity
$\rho_{xx}$ as a function of $B$ for linearly polarized radiation
and different polarization angles $\alpha$ regarding the transport
direction ($x$ direction). In b) curves are shifted for
clarity. In this scenario the damping parameter $\gamma > w$, then the $\rho_{xx}$ polarization immunity can be clearly
observed, especially for $B$ below the cyclotron resonance (see
vertical dashed line in the top panel). T=1.5K.
\newline

Figure 3 caption:
Calculated magnetoresistivity
$\rho_{xx}$ as a function of $B$ for linearly polarized radiation
and polarization angles $\alpha=0º$ and $\alpha=90º$ regarding the transport
direction ($x$ direction). These results correspond to
a scenario of high quality sample ($\gamma < w$),  and lower temperature ($T=0.7K$).
We observe how the $\rho_{xx}$ polarization immunity can be clearly
altered in function of the angle.
\newline


\begin{thebibliography}{15}
\bibitem{ina0}
J. I\~narrea, G. Platero and C. Tejedor, Semicond. Sci. Tech. {\bf
9}, 515, (1994);J. I\~narrea, G. Platero, Phys. Rew. B, {\bf 51}, 5244, (1995);
J. I\~narrea, G. Platero, Europhys. Lett.,
{\bf 34}, 43, (1996); J. I\~narrea, G. Platero, Europhys Lett.,  {\bf 33}, 477, (1996);
J. I\~narrea, R. Aguado, G. Platero, Europhys
Lett.  {\bf 40}, 417, (1997)

\bibitem{mani}
 R. G. Mani, J. H. Smet, K. von Klitzing, V. Narayanamurti,
W. B. Johnson, and V. Umansky, Nature(London) \textbf{420}, 646
(2002); R. G. Mani, V. Narayanamurti, K. von Klitzing, J. H. Smet,
W. B. Johnson, and V. Umansky, Phys. Rev. B\textbf{69}, 161306
(2004)

\bibitem{zudov}
M. A. Zudov,
 R. R. Du, L. N. Pfeiffer, and K. W. West,
Phys.
Rev. Lett. \textbf{90}, 046807 (2003)
\bibitem{smet}
J. H. Smet, B. Gorshunov, C. Jiang, L.N. Pfeiffer, K.W. West, V. Umansky,
M. Dressel, R. Dressel,  R. Meisels, F. Kuchar and K, von Klitzing ,  Phys. Rev. Lett., {\bf 95}, 116804, (2005)

\bibitem{maniapl}
R. G. Mani, Appl. Phys. Lett., {\bf 85}, 4962 (2004);
Appl. Phys. Lett., {\bf 91}, 132103 (2007);
Appl. Phys. Lett., {\bf 102}, 102107 (2008)


\bibitem{manipolar}
R. G. Mani, A. N. Ramanayaka, and W. Wegscheider
Phys. Rev. B {\bf 84}, 085308 (2011)


\bibitem{jesuspolar}
 A. N. Ramanayaka,R. G. Mani, J. Inarrea and W. Wegscheider
Phys. Rev. B {\bf 85}, 205315 (2011)

\bibitem{ina2}
J. I\~narrea and G. Platero, Phys. Rev. Lett. {\bf 94} 016806,
(2005); J. I\~narrea and G. Platero, Phys. Rev. B {\bf 72} 193414
(2005);J. I\~narrea and G. Platero, Appl. Phys. Lett.,  {\bf 89},
052109, (2006);J. I\~narrea and G. Platero, Phys. Rev. B,  {\bf 76},
073311, (2007);  J. I\~narrea, Appl. Phys. Lett. {\bf 90}, 172118,
(2007)



\bibitem{girvin}
A.C. Durst, S. Sachdev, N. Read, S.M. Girvin, Phys. Rev. Lett.{\bf
91} 086803 (2003)

\bibitem{dietel}
C.Joas, J.Dietel and F. von Oppen, Phys. Rev. B {\bf 72}, 165323,
(2005)

\bibitem{lei}
X.L. Lei and  S.Y. Liu, Phys. Rev. Lett.{\bf 91}, 226805 (2003)

%\bibitem{ryzhii}
%Ryzhii et al, Sov. Phys. Semicond. 20, 1299, (1986)

\bibitem{rivera}
P.H. Rivera and P.A. Schulz, Phys. Rev. B {\bf 70} 075314 (2004)

\bibitem{shi}
Junren Shi and X.C. Xie, Phys. Rev. Lett. {\bf 91}, 086801 (2003)

\bibitem{islamov}
A A Bykov, A K Bakarov,  D R Islamov  and I Toropov, JETP
letters, {\bf 84} 391, (2006); A A Bykov, D R Islamov, D V Nomokonov and
A K Bakarov,  JETP LETTERS, {\bf 86}, 608, (2008)


\bibitem{kerner}
E H Kerner,  Can. J. Phys, {\bf 36}, 371,  (1958)





\bibitem{fock}
V Fock, Z. Phys. {\bf 47}, 466, (1928);  C G Darwin, Proc.
Cambridge Philos. Soc., {\bf 27}, 86,(1930)


\bibitem{ina3}
J. Inarrea and G. Platero, Appl. Phys. Lett. {\bf 89},
172114, (2006);
J. I\~narrea and G. Platero, Appl. Phys Lett. {\bf 93}, 062104,
(2008); J. I\~narrea and G. Platero, Phys. Rev. B,. {\bf 78},
193310,(2008);J. I\~narrea, Appl. Phys Lett. {\bf 92},
192113,(2008)

\bibitem{ridley}
Ridley B. K 1993 {\it Quantum Processes in Semiconductors} (Oxford
University Press)

\bibitem{mani2}R. G. Mani,J. H. Smet, K. von Klitzing, V. Narayanamurti, W. B. Johnson, and V. Umansky,
  Phys. Rev. Lett. \textbf{92}, 146801, (2004); R. G. Mani, J. H. Smet, K. von Klitzing, V. Narayanamurti, W. B. Johnson, and V. Umansky,
 Phys. Rev. B\textbf{69}, 193304, (2004)


\bibitem{ina5}
Jesus Inarrea, arXiv:1203.1217v1;
Jesus Inarrea, arXiv:1109.5946v1


\bibitem{grads}
I.S. Gradshteyn and I.M. Ryzhik, Tables of Integrals, Series and
Products. 6th ed. Academic Press. (2000)

%\bibitem{yanhua}
%Yanhua Dai, R.R. Du,  L.N. Pfeiffer and K.W. West, Phys. Rev. Lett.,
%{\bf 105}, 246802, (2010).





\end{thebibliography}
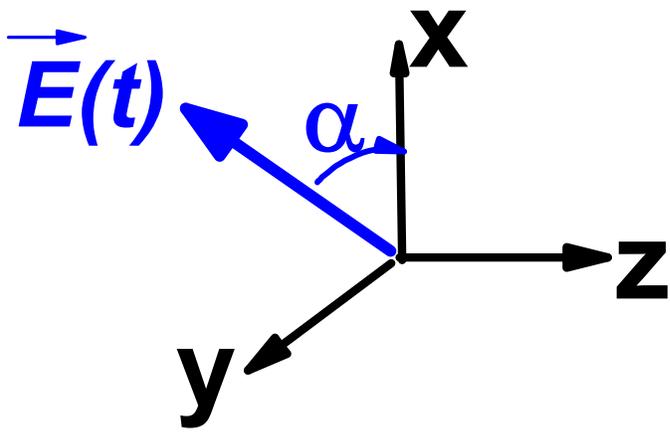
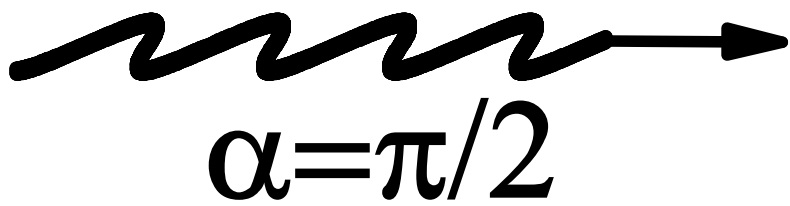
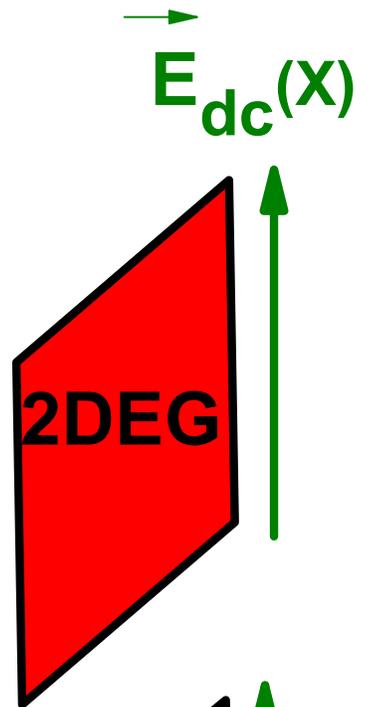
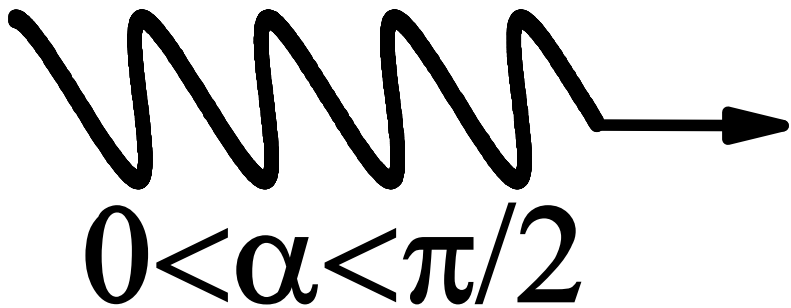
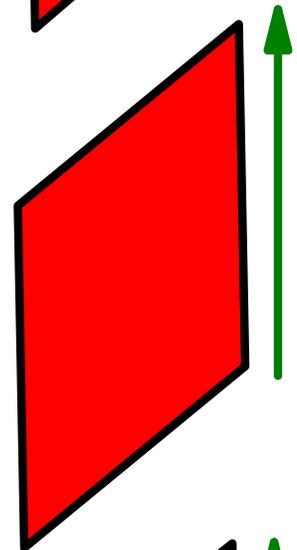
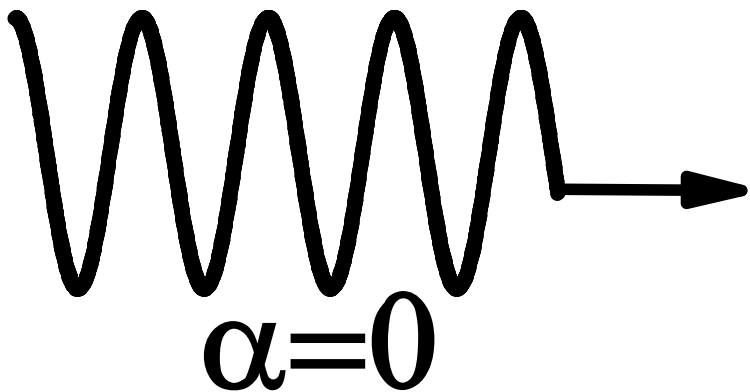
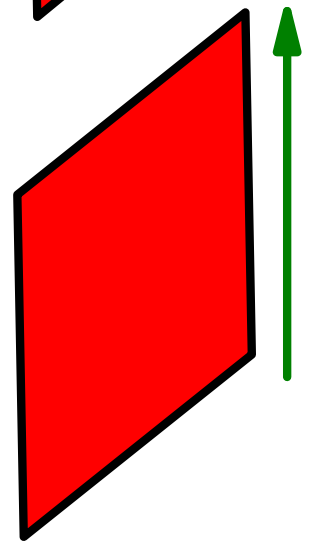
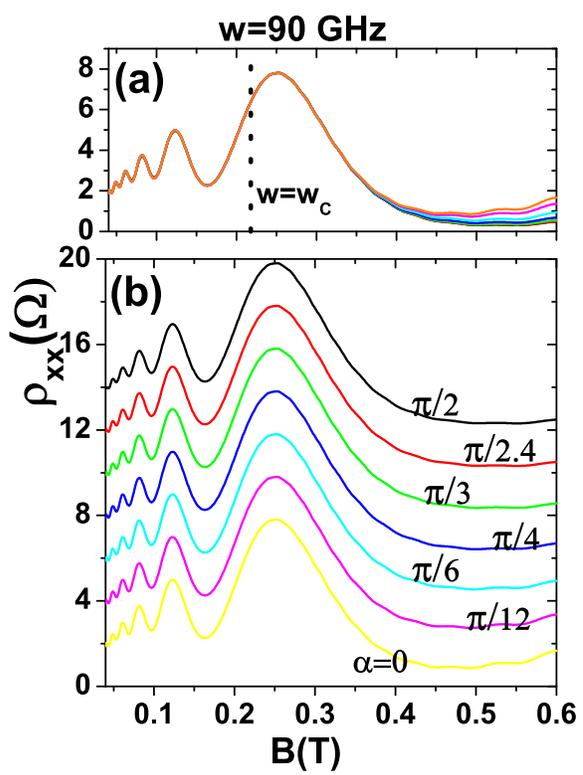
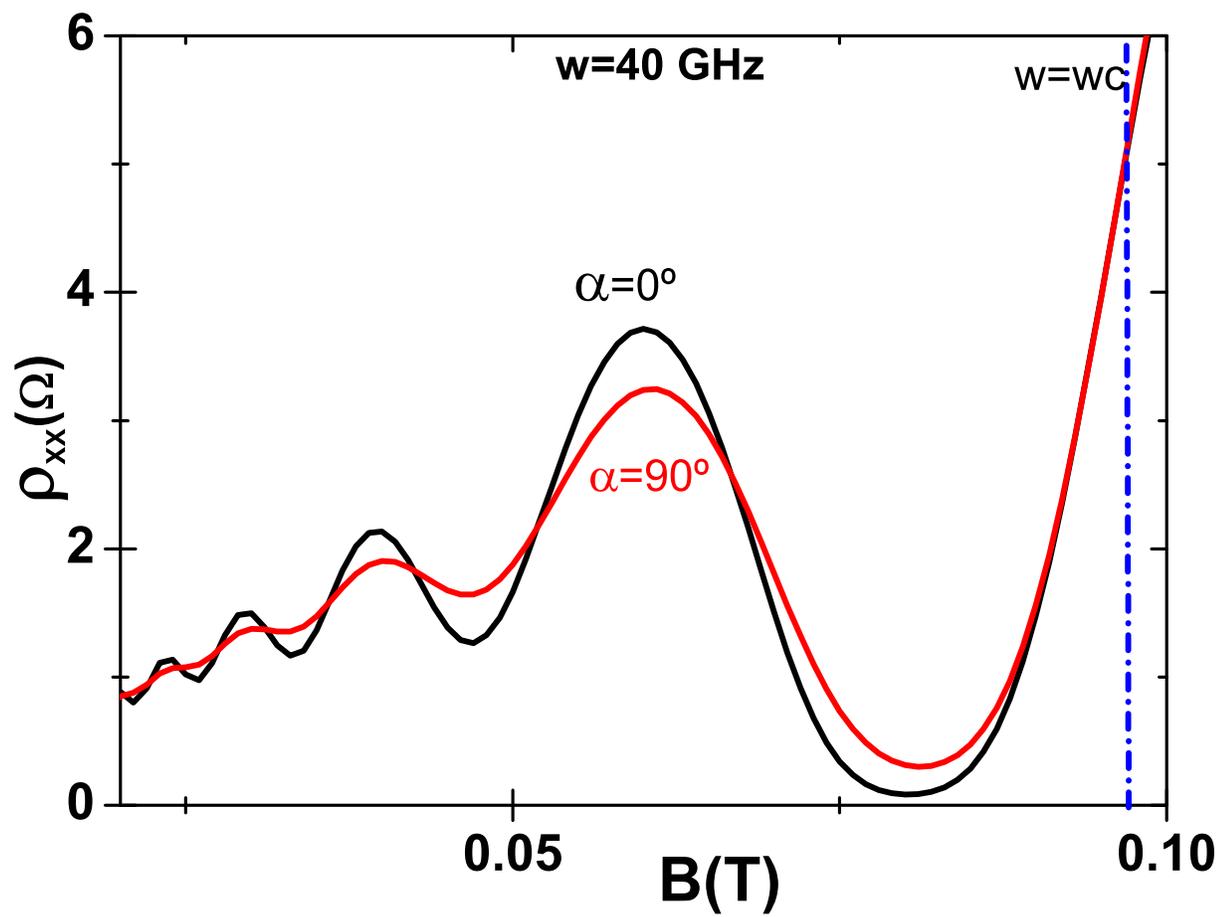
\end{document}